\documentclass{article}
\usepackage{spconf,amsmath,graphicx}
\usepackage{color}
\usepackage{amssymb}
\usepackage[ruled]{algorithm2e}
\usepackage{bm}
\usepackage{lipsum}

\usepackage{enumitem,hyperref}

\def\A{\bm{A}}

\def\I{\bm{I}}

\def\a{\mathbf{a}}

\def\s{\bm{s}}

\def\v{\bm{v}}

\def\x{\bm{x}}

\SetKwInput{kwEvaluate}{Evaluate}
\SetKwInput{kwSort}{Sort}
\SetKwInput{kwInput}{Input}
\SetKwInput{kwOutput}{Output}
\SetKwInput{kwInitialize}{Initialize}
\SetKwInput{kwParameters}{Params.}
\SetKwInput{kwDefaults}{Default init.}

\def\0{\mathbf{0}}
\def\1{\mathbf{1}}

\hypersetup{
    colorlinks=true,       % 
    linkcolor=red,        % 
    citecolor=blue,         % 
    filecolor=magenta,     % 
    urlcolor=cyan          % 
}

\title{(SP)\textsuperscript{2}-Net: 
A Neural Spatial Spectrum Method for DOA estimation}
\name{Lioz Berman,  Sharon Gannot, and Tom Tirer}
\address{Faculty of Engineering,
Bar-Ilan University, Ramat Gan, Israel \\
\{lioz.berman, sharon.gannot, tirer.tom\}@biu.ac.il}
\begin{document}
\maketitle
\begin{abstract}
We consider the problem of estimating the directions of arrival (DOAs) of multiple sources from a single snapshot of an antenna array, a task with many practical applications. % 
In such settings, the classical Bartlett beamformer is commonly used, as maximum likelihood estimation becomes impractical when the number of sources is unknown or large, and spectral methods based on the sample covariance are not applicable due to the lack of multiple snapshots. However, the accuracy and resolution of the Bartlett beamformer are fundamentally limited by the array aperture. 
In this paper, we propose a deep learning technique—comprising a novel architecture and training strategy—for generating a high-resolution spatial spectrum from a single snapshot. Specifically, we train a deep neural network that takes the measurements and a hypothesis angle as input and learns to output a score consistent with the capabilities of a much wider array. At inference time, a heatmap can be produced by scanning an arbitrary set of angles. % 
We demonstrate the advantages of our trained model, named (SP)\textsuperscript{2}-Net, over the Bartlett beamformer and sparsity-based DOA estimation methods.\footnote{Code can be found at \url{https://github.com/Liozb/SP2-Net}}
\end{abstract}

\section{Introduction}
\label{sec:intro}

Estimating the directions of arrival (DOAs) of radio frequency sources from the measurements of an antenna array is a fundamental problem encountered in radar and audio processing, communications, and more 
\cite{krim1996two}.
In many practical scenarios, such as those in automotive radar, the sources and/or receivers continuously move, and it is required to estimate the DOAs from a single realization (snapshot) of the measurements \cite{bialer2021super,bilik2019rise,sun2020mimo,tirer2020method}. 

When the number of sources is known, estimating the DOAs from a single snapshot can be based on the well-known
Maximum Likelihood (ML) approach.
For a single source, the ML estimate of the DOA is the angle of the peak of the classical Bartlett beamformer \cite{van2004optimum}. However, the computational complexity of the ML estimator grows significantly with the number of sources. For this reason, Bartlett beamforming is widely used in practice also for estimating multiple DOAs \cite{bialer2021super}, with the appealing property that the number of sources need not be specified before its application. Moreover, a user may prefer a method that produces a spatial spectrum rather than point estimates, e.g., when the results are passed to a higher-level decision-making block.
Yet, the Bartlett beamformer % 
suffers from bias when multiple sources are present and fails to resolve sources that are closer than the resolution limit defined by the array aperture.

Unfortunately, in the single snapshot case, many traditional spectral methods that offer improved resolution are not applicable. For example, MUSIC \cite{schmidt1986multiple} and ESPRIT \cite{roy1989esprit} cannot be used because the sample covariance cannot be decomposed into signal and noise subspaces without multiple snapshots.
Similarly, adaptive beamformers \cite{capon2005high} cannot be used as they are based on inverting the sample covariance \cite{ehrenberg2010sensitivity}. 
Nevertheless, better results than the Bartlett beamformer can still be obtained by sparsity-based methods, which formulate the problem as a sparse signal reconstruction task, and
estimate the DOAs from the peaks of the magnitude of the
recovered high-dimensional signal \cite{jeffs1998sparse, malioutov2005sparse,yang2018sparse}.

In recent years, there have also been attempts to estimate DOA using deep neural networks (DNNs) \cite{bialer2019performance}. Many methods frame the problem as a classification task, using single- or multi-label vectors as targets \cite{ma2022deep, de2022resnet, de2022deep, yang2022deepdoa}. However, these approaches are limited by a predefined angle grid, which restricts both the achievable accuracy and resolution. Other methods learn to enhance the covariance matrix and then apply subspace techniques for DOA estimation \cite{barthelme2021doa, wu2022gridless, shmuel2023deep}, but they require multiple snapshots. Some researchers have also explored end-to-end regression-based DOA estimation. Yet, this typically requires specifying the number of sources and does not provide a useful spatial spectrum \cite{bialer2019performance, xiong2022doa,xu2023end}.

\begin{figure*}
    \centering
    \includegraphics[width=0.9\linewidth]{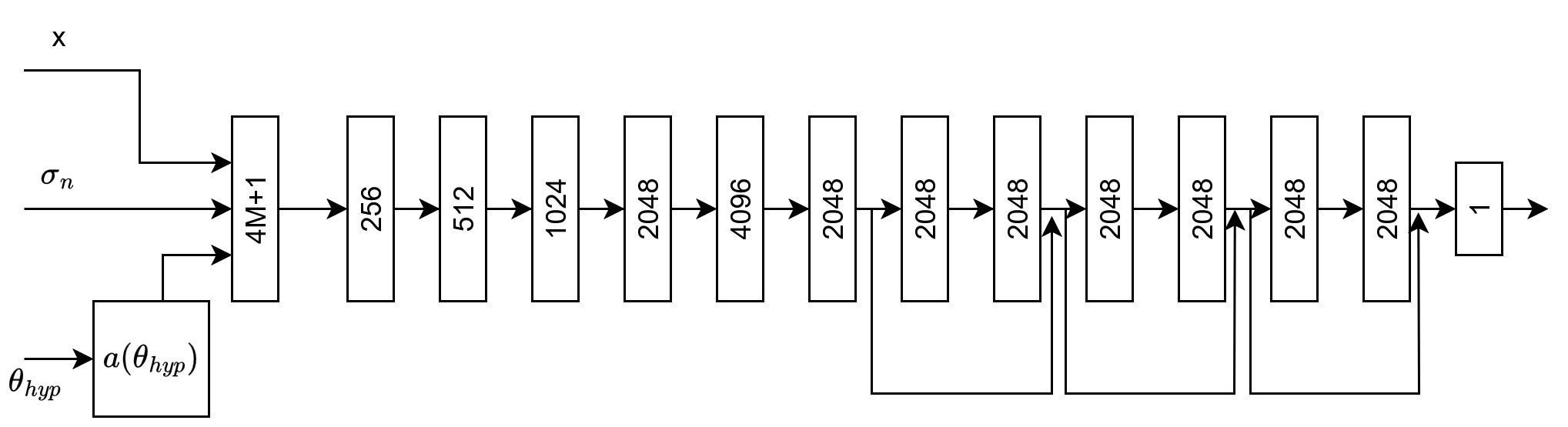}
    \vspace{-3mm}
    \caption{The proposed DNN architecture. The number of output neurons of each fully connected layer is indicated.}
    \label{fig:diagram}
    \vspace{-3mm}
\end{figure*}

In this paper, we propose the first deep learning-based method for generating a spatial spectrum for DOA estimation that is not restricted to a predefined grid of angles. Unlike traditional subspace methods, our approach operates on a single snapshot.
Inspired by research on coordinate-based networks \cite{tancik2020fourier,tancik2021learned}, we design a novel DNN architecture that receives a hypothesis angle as input and outputs a corresponding score. Unlike conventional coordinate-based networks, which mainly interpolate over the training measurements, our model also incorporates the observed array measurements and must generalize to unseen ones at test time, making the learning task considerably more challenging.
Our training strategy is also novel: the model is trained to output target scores corresponding to a spatial spectrum carefully designed to reflect the capabilities of a much larger array. We train a single model using array measurements with one to three active sources across a wide range of SNRs.

We name the resulting model (SP)\textsuperscript{2}-Net (abbreviating SPatial SPectrum), and benchmark its spatial spectra and statistical DOA estimation performance against the Bartlett beamformer and sparsity-based methods. Experiments demonstrate the advantages of (SP)\textsuperscript{2}-Net, particularly when the sources are closely spaced.

\section{Problem formulation}
\label{sec:format}

Consider an array of $M$ sensors receiving a single snapshot of narrowband signals from $Q$ sources under the far-field regime. % 
The sensors' measurements can be expressed as:
\begin{equation}
\label{eq:observation_model}
\x = \sum_{q=1}^{Q}\a(\theta_q)s_q + \v,
\end{equation}
where $\v \sim \mathcal{CN}(\0,\sigma_v^2\I_M)$ is an additive white circularly-symmetric complex-normal noise, $q$ is the source index, $s_q \in \mathbb{C}$ and $\theta_q$ are the unknown signal and direction of the $q$-th source, respectively, and $\a(\theta) \in \mathbb{C}^M$ denotes the array response to a signal that impinges the array from angle $\theta$.
Specifically, for a linear array, it is given by: {\small
\begin{equation}
\label{eq:steering_vec}
\a(\theta) = \frac{1}{\sqrt{M}}\left [ e^{j\frac{2pi}{\lambda}p_1 \cos(\theta)} , e^{j\frac{2pi}{\lambda}p_2 \cos(\theta)} , \dots , e^{j\frac{2pi}{\lambda}p_M \cos(\theta)}  \right ]^\top,
\end{equation}}
where $\lambda$ denotes the wavelength of the signal,
$(p_1,\dots,p_M)$ are the locations of the sensors, with the reference point in the array center, and the angle $\theta$ is measured with respect to the array axis (i.e., the boresight angle is $90^\circ$).

The problem at hand is to estimate the set of DOAs $\{\theta_q\}_{q=1}^Q$ from the array's measurements $\x$ given in \eqref{eq:observation_model}. 
For simplicity, we will present our approach for a Uniform Linear Array (ULA) of $M=16$ elements with half-wavelength $\lambda/2$ spacing.
However, the proposed approach can also be applied to other array configurations.

\section{The Proposed Method}
\label{sec:method}
In this section, we present our novel deep learning approach for spectrum-based DOA estimation from a single snapshot.
We discuss the proposed network architecture and training procedure separately in Sections \ref{sec:method_net} and \ref{sec:method_training}.

\subsection{Network Architecture}
\label{sec:method_net}

The key characteristic of the proposed DNN architecture is that it takes a hypothesis angle $\theta_{\text{hyp}}$ as input and outputs a scalar score. At inference, a spatial spectrum heatmap can be generated by scanning any arbitrary set of angles, chosen by the user. This flexibility distinguishes our method from most DNN-based {DOA} estimation approaches, which are typically trained to output a score vector over a fixed grid of angles.% 

This architectural choice is inspired by the research on coordinate-based networks \cite{tancik2020fourier,tancik2021learned}. However, the DNNs in this line of research focus on interpolating dense training samples (mostly belonging to a single image or a 3D object) and are not expected to generalize to other images/objects.
In contrast, our model also incorporates the measurements as input and must generalize to unseen ones at test time. A key component of coordinate-based networks is the \emph{positional encoding}, which transforms a low-dimensional coordinate into a higher-dimensional vector—a process shown to facilitate learning \cite{tancik2020fourier}. Whereas existing coordinate-based networks typically rely on heuristic choices for this transformation, in our case a natural encoding arises: the array manifold vector $\a(\cdot)$ in \eqref{eq:steering_vec}. This vector inherently captures the relation between a candidate angle and the array geometry, yielding a representation in the same ``feature space'' as the measurements.

Following the above discussion, the proposed DNN architecture takes as input both the measurement vector $\x \in \mathbb{C}^M$ and the steering vector corresponding to the hypothesis angle, $\a(\theta_{\text{hyp}}) \in \mathbb{C}^M$. In addition, we provide the standard deviation of the noise, $\sigma_v$, as input. This parameter can typically be estimated from time segments with no active sources, reflecting sensor-related noise characteristics. Moreover, $\sigma_v$ is a common input in existing {DOA} estimators; for example, in sparsity-based methods it is used to set hyperparameters such as the bound on a data-fitting constraint.

We now turn to the construction of the network. To operate with real-valued inputs only, we separate the real and imaginary parts of $\x$ and $\a(\theta_{\text{hyp}})$ into two distinct real components. This yields the input vectors $\tilde{\x} \in \mathbb{R}^{2M}$ and $\tilde{\a}(\theta_{\text{hyp}}) \in \mathbb{R}^{2M}$.
To handle different noise levels with a single model, $\sigma_v$ is concatenated to the input vector. 
Our architecture is a feed-forward network with fully connected layers,\footnote{We observed that fully connected layers yield better performance than convolutional layers, similar to the observation in \cite{bialer2019performance}.} ReLU activation functions, except for a sigmoid function at the output layer, and several skip connections, as illustrated in Fig.~\ref{fig:diagram}.
Specifically, the first learnable layer transforms the $(4M+1)$-dimensional input vector, which has a size of 65 for the examined setting of ULA with 16 elements, into a vector of size 256.
Then, the fully connected layers gradually increase the dimension of the hidden neuron layers until reaching a dimension of 2048, which is then kept fixed until the last learnable layer that outputs a scalar.
To facilitate optimization, we add skip connections between layers of size 2048.
In the sequel, we discuss how we train this network.

\subsection{Training Strategy}
\label{sec:method_training}

The main contribution of our learning strategy lies in the generation of the target score for each input tuple $(\theta_{\text{hyp}}, \x, \sigma_v)$. For clarity of presentation, we first describe the input samples used for training.

Following the well-established measurement model \eqref{eq:observation_model}, and consistent with prior work, we train the DNN using samples of $\x$ generated on-the-fly rather than relying on a fixed training set. The number of sources is drawn uniformly as $Q \in {1,2,3}$, with their true angles ${\theta_1, \ldots, \theta_Q}$ sampled uniformly within a 90\textdegree{} field of view (FOV), spanning 45\textdegree{}–135\textdegree{} around the array boresight. The first source signal $s_1$ is assigned unit magnitude, while the magnitudes of ${s_2, \ldots, s_Q}$ are drawn uniformly from $[0.5,1]$. The phases of all signals are drawn uniformly from $[0,2\pi]$.
The noise level $\sigma_v$ is determined from an SNR value (in dB) drawn uniformly from $[0,40]$ in steps of 1 dB, relative to the unit magnitude of the first source. This procedure guarantees that the SNRs of all other sources also fall within the range $[0,40]$ dB.

We now turn to our target score formulation, which also influences the distribution of the hypothesis angle $\theta_{\text{hyp}}$ used during training. Setting the target to 1 only when $\theta_{\text{hyp}}=\theta_q$ and to 0 otherwise is unsuitable, since (i) the probability of sampling $\theta_{\text{hyp}}$ exactly equal to a true {DOA} is zero, and (ii) we seek a spatial spectrum with meaningful physical interpretation. We therefore propose a different strategy.

Our approach builds on the well-established fact that the quality (in terms of accuracy and resolution) of the Bartlett beamformer improves substantially with increasing array aperture. Accordingly, we design the target score to emulate the performance of Bartlett beamforming on a much wider array, while further refining the spectrum by exploiting the availability of the true $\{\theta_q\}$ during training.

Our first step is to consider a virtual target array with more elements that handles the same source configuration.
Specifically, for the examined setting of ULA with $M=16$ elements and steering vector $\a(\theta) \in \mathbb{C}^{16}$, we choose to use a virtual target ULA with $M_{\text{tg}}=4M=64$ elements and denote its steering vector by $\a_{\text{tg}}(\theta) \in \mathbb{C}^{64}$. 
In principle, one may use a higher value of $M_{\text{tg}}$. However, note that as the virtual aperture increases, the resulting Bartlett main lobe becomes narrower, which makes the learning task more challenging since most of the FOV corresponds to near-zero scores.

We now present two modifications to standard Bartlett beamforming, applied during training to enhance the quality of the target spectrum.  
First, we define the target array to be noise-free, i.e., disregarding the value of $\sigma_v$, which yields the spectrum 
$ \left| \a_{\text{tg}}^{\mathrm{H}}(\theta_{\text{hyp}}) \sum_{q=1}^{Q} \a_{\text{tg}}(\theta_q) s_q \right|^2 $.  
Second, although the Bartlett beamformer is optimal in the single-source case, it exhibits bias in the presence of multiple sources and is further affected by the varying magnitudes of $\{s_q\}$. To avoid these limitations, we instead define the training target score as the \emph{maximum} of fixed-magnitude single-source spectra:  
\begin{equation}
\label{eq:target}
t(\theta_{\text{hyp}},\{\theta_q\}) = \max_{q=1,\ldots,Q} 
\left| \a_{\text{tg}}^{\mathrm{H}}(\theta_{\text{hyp}})\a_{\text{tg}}(\theta_q) \right|^2 .
\end{equation}

For each measurement $\x$ and each DOA $\theta_q$ used to generate $\x$, we sample $K=80$ hypothesis angles while ensuring good coverage of the angular range where the main lobe of the virtual target beamformer exhibits its 3~dB decay. This range is given by $[\theta_q-\tfrac{\Delta}{2},\,\theta_q+\tfrac{\Delta}{2}]$ with $\Delta=\tfrac{2}{M_{\text{tg}}}\tfrac{180}{\pi}$ (note that $\tfrac{2}{M_{\text{tg}}}$ equals $\lambda$ divided by the virtual aperture). Specifically, we uniformly sample $0.5K$ hypothesis angles within this range and the remaining $0.5K$ outside it.

The loss function is the mean squared error (MSE) between the network’s predicted scores and the target values stated in \eqref{eq:target}.
As we cover the SNR range of $[0,40]$ dB, to prevent the loss from being dominated by updates from low-SNR samples, we weight the loss function according to the SNR, using its linear scale, normalized to a maximal value of 1. % 
We use the ADAM optimizer with a learning rate of $\alpha=0.001$ and 200 training scenarios (i.e., different measurements) per optimization iteration.
We use a fixed validation set, generated similarly to the training set, to track the optimization process and stop it at convergence.

\begin{figure}[t]
    \centering
    \includegraphics[width=0.7\linewidth, trim={10 10 0 30}, clip]{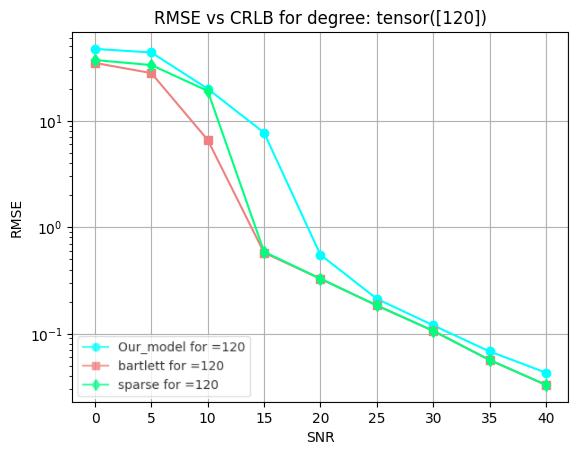}
    \vspace{-3mm}
    \caption{RMSE [deg] vs. SNR [dB] for a source at 120\textdegree{}.}
    \label{fig:single_source}
\end{figure}

\vspace{-1mm}

\section{Experiments}

\vspace{-1mm}

\begin{figure}[t]
    \centering
    \includegraphics[width=0.7\linewidth, trim={10 10 0 30}, clip]{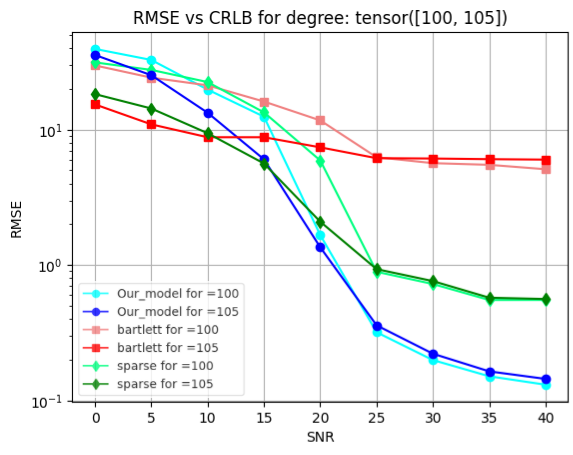}
    \includegraphics[width=0.32\linewidth]{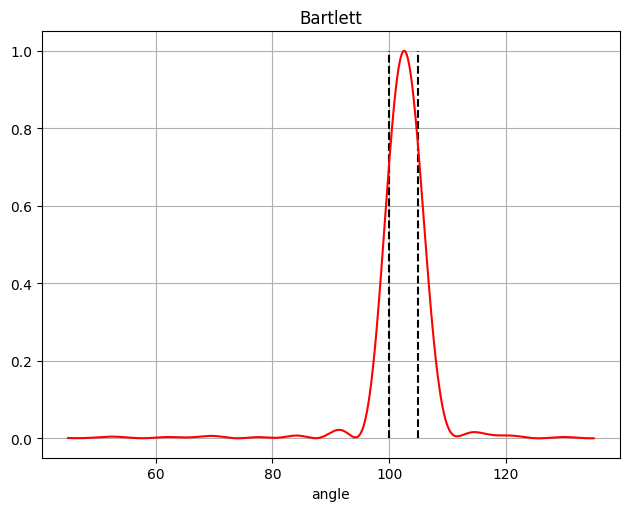}
    \includegraphics[width=0.32\linewidth]{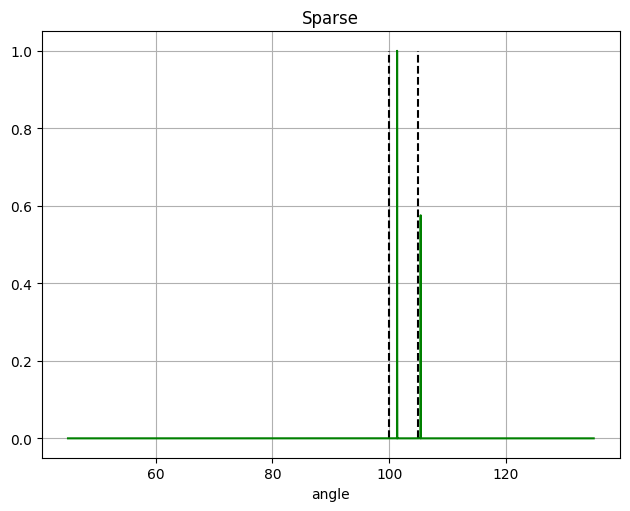}
    \includegraphics[width=0.32\linewidth]{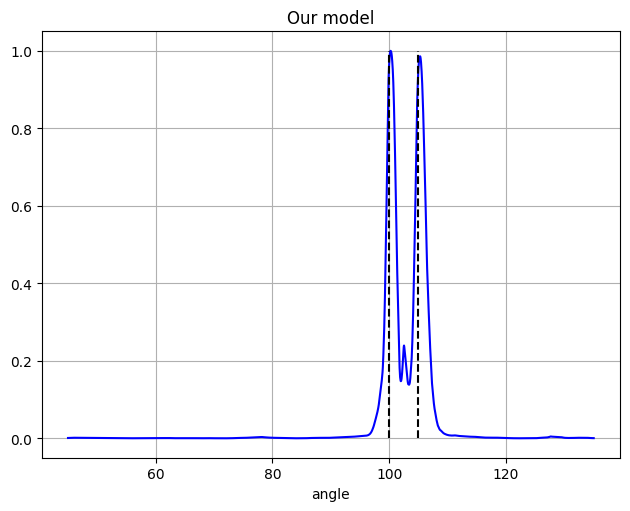}
    \vspace{-3mm}
    \caption{Sources at 100\textdegree{} and 105\textdegree{}. Top: RMSE [deg] vs. SNR [dB]; Bottom: Spectra for a realization at SNR 25 dB.}
    \label{fig:2 sources}
    \vspace{-2mm}
\end{figure}

\begin{figure}
    \centering
    \includegraphics[width=0.7\linewidth, trim={10 10 0 30}, clip]{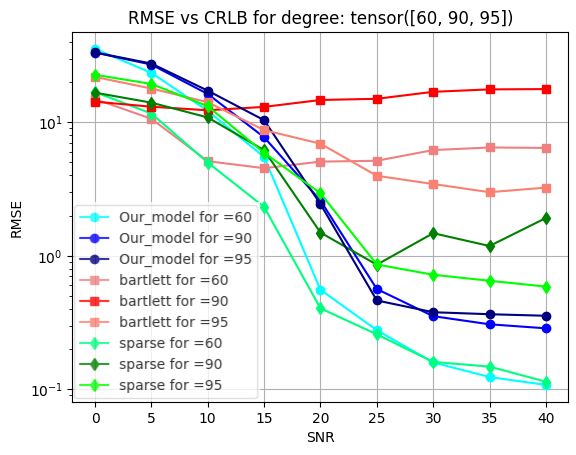}
    \includegraphics[width=0.32\linewidth]{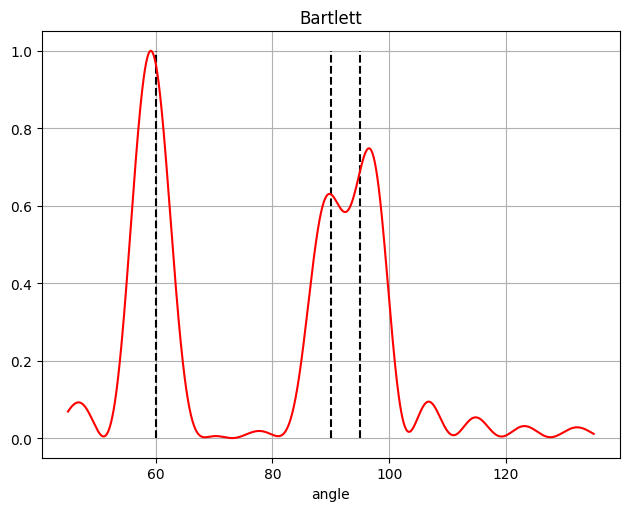}
    \includegraphics[width=0.32\linewidth]{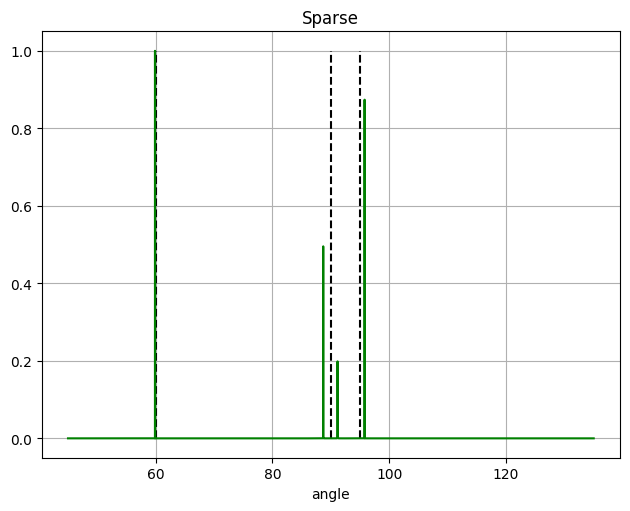}
    \includegraphics[width=0.32\linewidth]{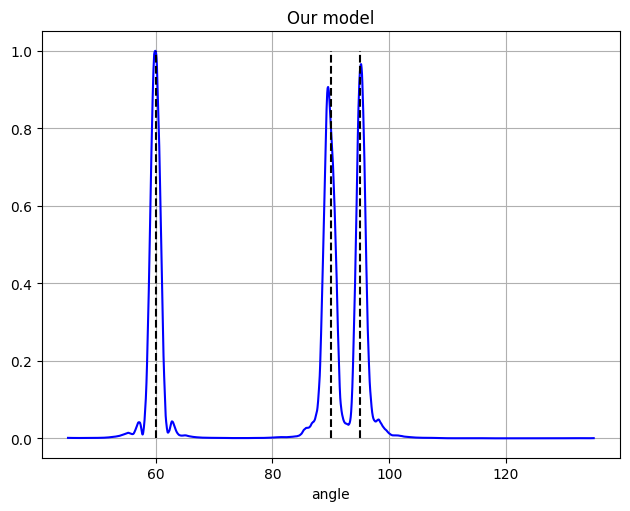}
    \vspace{-3mm}
    \caption{Sources at 60\textdegree{}, 90\textdegree{} and 95\textdegree{}. Top: RMSE [deg] vs. SNR [dB]; Bottom: Spectra for a realization at SNR 30 dB.}
    \label{fig:3 sources}
    \vspace{-3mm}
\end{figure}

\begin{figure}[t]
    \centering
    \includegraphics[width=0.32\linewidth]{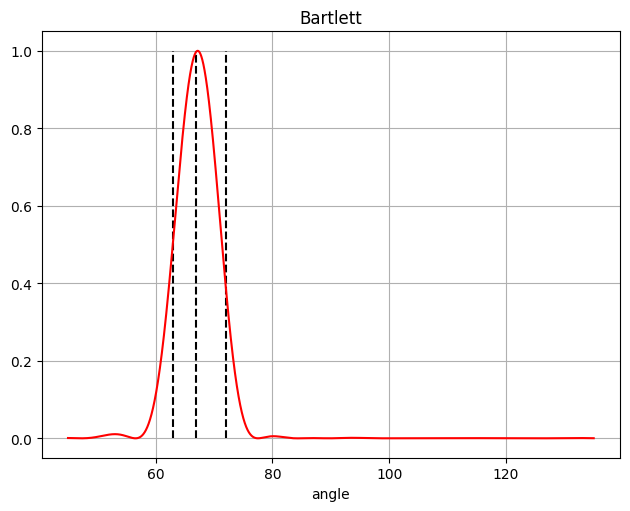}
    \includegraphics[width=0.32\linewidth]{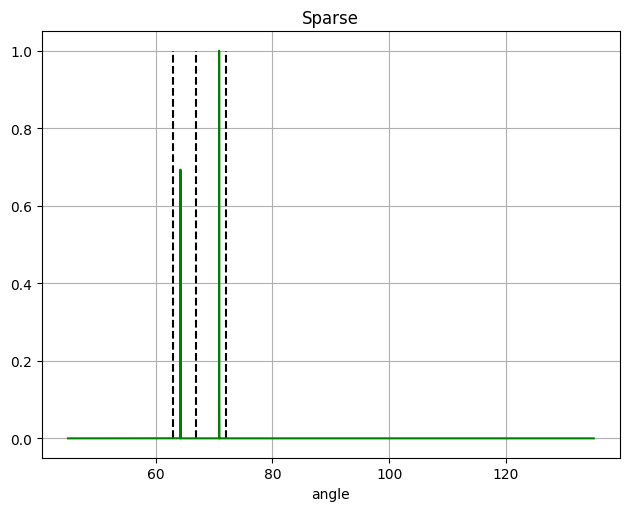}
    \includegraphics[width=0.32\linewidth]{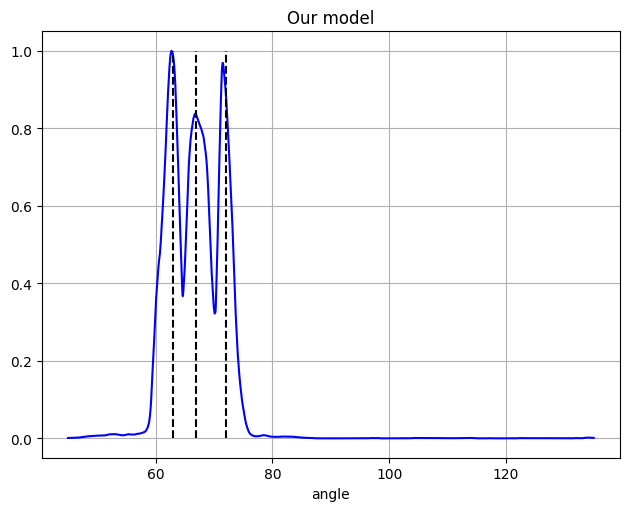}
    \vspace{-3mm}
    \caption{Spectra of a realization at SNR 30 dB for sources at 63\textdegree{}, 67\textdegree{} and 72\textdegree{}.}
    \label{fig:3 sources dense spectra}
    \vspace{-3mm}
\end{figure}

Consider a setup with a single snapshot from a ULA with $M=16$ elements, as exemplified in the paper. Recall, though, that the proposed approach is not restricted to ULAs.

We compare the performance of our trained network, dubbed (SP)\textsuperscript{2}-Net, with two reference methods: (i) the Bartlett beamformer and (ii) the widely-used sparsity approach \cite{malioutov2005sparse,yang2018sparse,tirer2022direction}. The methods are compared in terms of the root mean square error (RMSE) of their DOAs estimation, computed over 500 Monte-Carlo experiments at each SNR between 0 dB and 40 dB, as well as by examining their output spectra. % 

For all methods, the output spectra are evaluated over a discrete set of angles $\{\theta_i\}_{i=1}^{N_{\text{hyp}}}$ covering the 90\textdegree{} FOV with a resolution of 0.01\textdegree{}.
For each method, estimating $Q$ DOAs involves searching for the $Q$ highest peaks (local maxima) in the spectrum. 
The sparsity-based method solves the convex problem:
$
\min_{\tilde{\s}\in \mathbb{C}^{N_{\text{hyp}}}} \|\tilde{\s}\|_1 \,\,\, \text{subject to} \,\,\, \|\x - \A\tilde{\s}\|^2 \leq C M \sigma_v^2
$,
where $\A=[\a(\theta_1),...,\a(\theta_{N_{\text{hyp}}})]$ is the array manifold matrix, and $C$ is a hyperparameter. We found that $C=2$ provides the best results, similarly to \cite{tirer2022direction}. 

We begin with the single-source scenario.
Fig.~\ref{fig:single_source} shows the RMSE [deg] versus SNR [dB] for all methods with a source located at $120^\circ$. In this case, the Bartlett beamformer is equivalent to ML estimation and is therefore optimal at sufficiently high SNR. The sparsity-based method achieves performance comparable to that of the Bartlett beamformer, while our model falls slightly short, although the gap is minor.

We next examine more challenging multi-source scenarios, where the Bartlett beamformer is no longer optimal.
Figure~\ref{fig:2 sources} presents the results for $Q=2$ sources located at $100^\circ$ and $105^\circ$, while Fig.~\ref{fig:3 sources} depicts the case of $Q=3$ sources at $60^\circ$, $90^\circ$, and $95^\circ$. In both scenarios, the sources have equal magnitudes and random phases.
To further compare the methods, each figure also includes a visualization of the spectra for a representative realization. An additional spectrum visualization is provided in Fig.~\ref{fig:3 sources dense spectra} for $Q=3$ sources at $63^\circ$, $67^\circ$, and $72^\circ$.
These figures clearly demonstrate that our (SP)\textsuperscript{2}-Net outperforms the competing methods.
In particular, Figs.~\ref{fig:2 sources} and \ref{fig:3 sources} highlight the poor RMSE performance of the Bartlett beamformer. The sparsity-based method improves upon Bartlett but still falls short of the accuracy achieved by (SP)\textsuperscript{2}-Net, particularly for closely spaced sources. The spectral visualizations further reveal the larger bias of the competing methods and their difficulty in reliably identifying all closely spaced sources. 

\vspace{-2mm}

\section{Conclusion}

\vspace{-1mm}

In this work, we proposed (SP)\textsuperscript{2}-Net, the first DNN-based method, to the best of our knowledge, for generating spatial spectra over arbitrary hypothesis angles from a single snapshot. We introduced novel ideas in both architecture and training. Comparative experiments with standard methods demonstrated the clear advantages of the proposed method in estimating the DOAs of closely spaced sources.
% 
%Our code and trained model will be made available upon acceptance. 
% 

% 

% 

% 
\bibliographystyle{IEEEbib}
\bibliography{strings,refs}

\end{document}